\documentclass[prb,preprintnumbers,amsmath,amssymb,twocolumn,superscriptaddress]{revtex4-1}

\usepackage[dvips]{graphicx}
\usepackage{dcolumn}
\usepackage{bm}
\usepackage[dvips]{color}
\usepackage{natbib}

\usepackage{amsmath}	
\usepackage{amsfonts}

\begin{document}

\title{
Fulde-Ferrell state in ferromagnetic chiral superconductor with magnetic domain wall
}

\author{Yasuhiro Tada}
\email{tada@issp.u-tokyo.ac.jp}
\affiliation{Institute for Solid State Physics, The University of Tokyo,
Kashiwa 277-8581, Japan}

\begin{abstract}
Motivated by the recent theoretical and experimental progress in the heavy fermion system UCoGe,
we study ferromagnetic chiral superconductors in the presence of magnetic domains.
Within mean field approximations, it is shown that chiral superconducting domains
are naturally induced by the ferromagnetic domains.
The domain wall current flows in the opposite direction
to the naively expected one as in $^3$He-A phase due to contributions from ``unpaired electrons".
Consequently, the domain wall current flows in the same direction with that of surface currents
when the magnetic domain wall lies parallel to the sample surface, and therefore
they contribute to the net current along the whole sample.
We find that, due to the non-cancellation between the domain wall current and surface current,
a Fulde-Ferrell-like superconducting state can be stabilized in an anisotropic sample
for all the temperatures below the superconducting transition temperature.
\end{abstract}

\pacs{Valid PACS appear here}

\maketitle

\section{introduction}
Interplay of ferromagnetism (FM) and superconductivity (SC) has been a central issue
in the study of magnetic superconductors.
A FM order generally breaks Cooper pairing through Zeeman coupling and
Lorentz force, and therefore it is usually difficult to realize
coexistence of FM and SC orders. 
However, in a spin triplet pairing state,  weakness of Pauli depairing effect could allow
microscopic coexistence of the two seemingly antagonistic orders.
Indeed, it is expected that pseudo-spin triplet states in the presence of FM orders are realized
in the family of uranium based ferromagnetic superconductors UGe$_2$, URhGe, and UCoGe~\cite{pap:Saxena2000,pap:Aoki2001,pap:Huy2007}.

Among these uranium systems, UCoGe is particularly interesting since it has been proposed to be a topological superconductor.
In the previous theoretical study combined with the NMR experiments, 
it was demonstrated that the upper critical field $H_{c2}$ of UCoGe at ambient pressure is well explained based on the so-called 
A-state symmetry rather than the other group theoretical candidate B-state symmetry~\cite{pap:Hattori2012,pap:Hattori2014,pap:Tada2013,pap:Tada2016}. 
The $d$-vector of the superconducting A-state in the simplest pseudo-spin 1/2 basis
is of the form $d\sim (a_1k_x+ia_2 k_y, a_3 k_y+ia_4 k_x,0)$ with the real coefficients $\{a_j\}$~\cite{pap:Mineev2002}, 
and therefore UCoGe at ambient pressure in the A-state
can be considered as a chiral superconductor in the presence of ferromagnetic order~\cite{pap:SatoFujimoto2016}.
Besides, it is shown that there exist nodes in quasi-particle spectrum in the FM+SC state
which are protected by the nonsymmorphic magnetic space group, in addition to the possible Weyl points at poles~\cite{pap:Nomoto2016}.
The pairing symmetry in the paramagnetic state at high pressures is also proposed to be topological~\cite{pap:CheungRaghu2016}.
In such a topological magnetic superconductor, a natural question is that how the FM and topological character of the SC
are correlated.

In addition to the momentum space topology in a ferromagnetic topological superconductor, 
FM degrees of freedom gives rise to real space non-trivial structure defined
by magnetic domains in the ordered states.
Experimentally, FM domains have been observed in UCoGe by scanning SQUID at ambient pressure, and the domain wall (DW) width was estimated to be
very small, $\sim 0.1$-1nm, as expected from the strong Ising anisotropy of UCoGe~\cite{pap:Hykel2014}.
The magnetic flux density is changed only slightly as the system enters the FM+SC phase from the non-superconducting FM phase.
Based on this results, it was concluded that Meissner effect is weak and the spontaneous vortex state without external magnetic field 
would be realized
in UCoGe, although vortices could not be directly seen because of the limited spatial resolution of the scanning SQUID.
Under applied pressure, however, one would also expect that a Meissner state could be realized,
since the magnetic moment would be suppressed so that it cannot stabilize the spontaneous vortex state.

In this study, motivated by the theoretical identification of chiral superconductivity and the experimental observation of magnetic domains
in UCoGe, we investigate a ferromagnetic chiral superconductor,
especially focusing on interplay between real space FM domains and momentum space topology of SC
when the system is in a Meissner phase.
In such a set up, one would expect that chiral SC domains are induced by FM domains since the magnetic flux density
arising from the FM order may stabilize one of the chiralities of Cooper pairing~\cite{pap:Mineev2002,pap:Sumiyoshi2014}.
Once chiral SC domains are formed, DW current due to the non-trivial momentum space topology will flow along the DW~\cite{book:Volovik2003}. 
Within mean field approximations, we show that chiral SC domains are indeed naturally induced and the spatial distribution of domains
simply coincides with that of the FM domains.
We further demonstrate that the direction of the DW current
is opposite to the naively expected one for a stable DW as in the case of $^3$He-A phase~\cite{pap:Tsutsumi2014, pap:Volovik2015}.
We propose that, as a result of the non-cancelation between the DW and surface currents, 
a Fulde-Ferrell (FF) like stripe superconducting state~\cite{pap:FF1964, pap:LO1965}
 can be stabilized in an anisotropic sample
of a ferromagnetic chiral superconductor in the Meissner state in the presence of FM domain walls.
The microscopic physical origin of the FF-like state is also discussed in terms of hidden normal (non-superconducting) state components, ``unpaired electrons", 
which are an important key
for understanding paired states~\cite{pap:Tada2015PRL,pap:Prem2017}.

\section{model}
In this study, we consider a simple model of ferromagnetic chiral superconductors. 
Although we take UCoGe as a prototypical example of ferromagnetic chiral superconductors and use a specific model which is relevant to UCoGe,
we will discuss their general properties based on the concrete model.
Our results are basically applicable also to other chiral superconductors/superfluids such as $^3$He-A phase
in confined geometry with some modifications, if one simply assumes that chiral SC domains are created by some reasons.
While it is essentially difficult to control and identify SC domains in those systems, 
they can be naturally induced by FM domains in ferromagnetic chiral superconductors as will be discussed later.

The on-site level scheme of UCoGe due to the spin-orbit interaction and crystal electric field
has not been identified experimentally, and the calculated band structure looks very complicated~\cite{pap:UCoGe_band}.
However, the resistivity shows rather simple behaviors:
it is nearly isotropic and exhibits a single coherence peak around $T\sim$30-40K for example in the $c$-direction~\cite{pap:Hattori2012,pap:Hattori2014}.
This would suggest that UCoGe could be effectively described by a pseudo-spin 1/2 fermions
with a nearly isotropic Fermi surface.
In the present study, such fermions are simply denoted as spin 1/2 electrons for simplicity.
Here we introduce a simple model for effectively describing superconductivity in UCoGe,
\begin{align}
H=-\sum_{\langle ij\rangle\sigma}t_{ij}c^{\dagger}_{i\sigma}c_{j\sigma}-\sum_ih^{\rm ex}_iS^z_i
-\sum_{\langle ij\rangle\sigma}\frac{g_{ij}}{4}n_{i\sigma}n_{j\sigma},
\label{eq:H}
\end{align}
where $S^z_i=\sum_{ss'}c^{\dagger}_{is}\sigma^z_{ss'}c_{is'}/2$ and $n_{i\sigma}=c^{\dagger}_{i\sigma}c_{i\sigma}$.
For simplicity, we consider a layered square lattice with equal lattice spacing for all the directions
and $i,j$ represent the site positions.
Since the $z$-direction degrees of freedom is not important for our study, we suppress the $z$-index. 
The first term is hopping between the nearest neighbor sites only within a two-dimensional $xy$-plane
 and also includes chemical potential $t_{ii}=\mu$. Introduction of $z$-direction hopping does not change our main results.
The hopping integral includes 
a classical vector potential $t_{ij}=t\exp(i\tilde{A}_{ij})$ where $\tilde{A}_{ij}=(e/c\hbar)A((i+j)/2)\cdot(i-j)$
with the electron charge $e$.
The vector potential is simply determined by Maxwell equation with a suitable boundary condition. 
Gauge fluctuations are completely neglected in this study for which conventional BCS theory works well,
although inclusion of gague fluctuations would modify the standard BCS description~\cite{pap:TadaKoma1,pap:TadaKoma2}.
The second term is the exchange term in the FM state which may arise from effective interactions
such as $S^z_iS^z_j$ but is a given function in the present study.
The last term describes an effective interaction which leads to chiral SC corresponding to the A-state of UCoGe.
The interaction is non-zero only for the nearest neighbor sites in the same $xy$-plane, $g_{i,i\pm\hat{x}}=g_{i,i\pm\hat{y}}=g$ and
$g_{i,i\pm\hat{z}}=0$, and acts only on the electrons with the same spins.
The resulting chiral SC under these assumptions is consistent with the previous theory where the electrons interact 
through Ising magnetic fluctuations as clarified experimentally~\cite{pap:Hattori2012,pap:Hattori2014,pap:Tada2013,pap:Tada2016}.

To describe a FM domain structure, we consider a simple functional form of $h^{\rm ex}_i$,
\begin{align}
h^{\rm ex}_i=h^{\rm ex}_0\tanh((x-x_{DW})/d_{DW}),
\end{align}
for which the DW is on the $yz$-plane located at $x=x_{DW}$ with the width $d_{DW}$.
The position $x_{DW}$ is assumed to be the center of the system in the $x$-direction.
Since UCoGe has strong Ising anisotropy, the DW could be well described by the above
function. Experimentally, the DW width in UCoGe is estimated as $d_{DW}\sim 0.1$-1 nm,
and typical domain size is $\sim 1$-10 $\mu$m~\cite{pap:Hykel2014}.
It is noted that $h^{\rm ex}_i$ should be regarded as a renormalized quantity including
Zeeman coupling with the self-induced magnetic flux density which is expected to be small compared to 
an effective exchange coupling between spins.
In the SC state, the magnetic flux density distribution would be modified from that in the non-superconducting state,
which would also affect FM structure and $h^{\rm ex}_i$ in general.
Although such an effect could be studied by a full self-consistent calculation for both FM and SC orders with 
using realistic model parameters (such as $\varepsilon_{CR}$ defined later and effective magnetic interactions between spins), 
we simply fix $h^{\rm ex}_i$ as a given parameter in this study.
We note that, 
although the above FM domain structure may be oversimplified, it is important to understand the simplest case
in order to develop further theoretical understanding.
As a straightforward generalization, we will briefly discuss multi-domain cases at the end of Sec.III.
Similar FM domain structures have been studied for general FM superconductors also in different contexts
~\cite{pap:Buzin2003_1,pap:Buzin2003_2,pap:Buzin2005,pap:Yang2004,pap:Lorenz2005}.

\begin{figure}[htbp]
\begin{center}
\includegraphics[width=0.8\hsize,height=0.3\hsize]{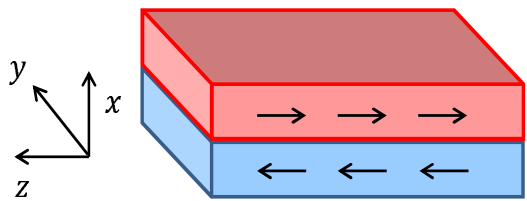}
\end{center}
\caption{The anisotropic system with $L_x\ll L_y, L_z$. The FM domain wall on $yz$-plane seperates the positive magnetization domain and 
negative magnetization domain. }
\label{fig:film}
\end{figure}

We consider an anisotropic system with an open boundary condition for $x$-direction and
periodic boundary conditions for $y,z$-directions as shown in Fig.~\ref{fig:film}.
The system size is $L_x\times L_y\times L_z$ with $L_x\ll L_y, L_z$, and is fixed as $L_x=200a$ and $L_y=2000a$ in the present study.
Although the $z$-index is redundant, $L_z$ is assumed to be large enough so that 
only the $z$-component of the magnetic flux density is non-zero
when a charge current flows in the $xy$-plane.
In addition,
although the spontaneous vortex state may be realized in UCoGe at ambient pressure, 
we rather focus on Meissner states which would be stable under pressure where magnetization and corresponding
magnetic flux density are suppressed.
The spontaneous vortex state is left for future studies.
We note that
our results are qualitatively robust against the thickness $L_x$ as long as the system shape is anisotropic,
although they depend quantitatively on the thickness as will be discussed later.
As was mentioned before, our results are rather general for ferromagnetic chiral superconductors,
although we use the specific model Eq.\eqref{eq:H} which is relevant to UCoGe.

For a given vector potential, we use a mean field approximation for 
the conventional order parameter
\begin{align}
g\langle c_{i\sigma}c_{i+\hat{x},\sigma}\rangle&=-i\Delta^{x}_{i\sigma}, \nonumber \\
g\langle c_{i\sigma}c_{i+\hat{y},\sigma}\rangle&=\Delta^{y}_{i\sigma}.
\end{align}
In the presence of translational symmetry in the Meissner state,
$\Delta^{x/y}_{i\sigma}$ depend only on $x$ and the mean field Hamiltonian reads
\begin{align}
H_{\rm MF}&=1/2\sum_{xx',k_{y}}C^{\dagger}_{xk_{y}}{\mathcal H}_{{\rm BdG},xx'}(k_{y})C_{x'k_{y}}
-\sum_{\langle ij\rangle}t_{ij}+E_{\rm c},
\label{eq:H_MF}
\end{align}
where $C_{xk_{y}}=(c_{xk_{y}\uparrow},c_{xk_{y}\downarrow},c^{\dagger}_{x,-k_{y},\uparrow},c^{\dagger}_{x,-k_{y},\downarrow})$
and ${\mathcal H}_{\rm BdG}$ is the Bogoliubov-de Gennes (BdG) Hamiltonian.
$E_{\rm c}=\sum_{\langle ij\rangle \sigma}g_{ij}/4|\langle c_{i\sigma}c_{j\sigma}\rangle|^2$ is the condensation energy.
Then we compute lattice electric current 
\begin{align}
j_{i,y}&=j_{i\uparrow,y}+j_{i\downarrow,y}\nonumber \\
&=\frac{-ie}{\hbar} \sum_{\sigma} a[t_{i,i+\hat{y}}c^{\dagger}_{i\sigma}c_{i+\hat{y}\sigma}-t_{i+\hat{y},i}c^{\dagger}_{i+\hat{y}\sigma}c_{i\sigma}],
\end{align}
where $e<0$ is the electric charge and $a$ is the lattice constant.
Similarly, $z$-component of the spin current is defined as
\begin{align}
j_{i,y}^{\rm spin}&=j_{i\uparrow,y}-j_{i\downarrow,y}.
\end{align}
We choose a gauge among the Landau gauge $\tilde{\nabla}\cdot \tilde{A}=0$ so that $\tilde{A}_{i,i+\hat{x}}=\tilde{A}_{i,i+\hat{z}}=0$,
where $\tilde{\nabla}$ is the lattice dimensionless derivative.
The magnetic flux density for each plaquette $\tilde{B}_{iz}= (\tilde{\nabla}\times \tilde{A})_{iz}
=\tilde{A}_{i+\hat{x}+\hat{y},i+\hat{x}}-\tilde{A}_{i+\hat{y},i}$ is calculated by the lattice Maxwell equation
in a dimensionless form,
\begin{align}
(\tilde{\nabla}\times \tilde{B})_{iy}=4\pi \frac{at}{\phi_0^2}\tilde{j}^{\rm tot}_{i,y},
\end{align}
where $\phi_0=c\hbar/|e|$ and $\tilde{j}^{\rm tot}=\tilde{j}_{}+\tilde{\nabla}\times \tilde{M}^{\rm spin}$.
The lattice spacing for the $z$-direction has been taken to be the same value as those in $x, y$-directions, $a$.
We use a boundary condition on $\tilde{B}$ so that $\tilde{B}_{i,z}=0$ just outside the sample,
which corresponds to zero external magnetic field.
The dimensionless current and spin magnetization are defined as $\tilde{j}=j\hbar/(|e|at)$ and 
$\tilde{M}^{\rm spin}_{i,z}=-2\langle S^z_i\rangle$ where we have simply assumed $g_{\rm eff}=2$ for the effective $g$-factor.
This equation reduces to the conventional Maxwell equation when the lattice dimensionless derivative $\tilde{\nabla}$
is replaced by the derivative operator in the continuum multiplied by the lattice constant $a$.
The dimensionless numerical factor $\varepsilon_{CR}\equiv at/\phi_0^2$ is estimated to be $\sim 3\times 10^{-6}$
for $t=0.2$eV and $a=5$\AA.
When we use this value in numerical calculations, the resulting penetration depth $\lambda$ is found to be
of the order $100a=50$nm or even longer.
Since the effective hopping $t$ should be regarded as a renormalized quantity and be
smaller than $0.2$eV when considering UCoGe, 
one might obtain $\lambda\sim 1\mu$m as evaluated experimentally if a more realistic value of $t$ were used.
In the present study, however, we use a large $\varepsilon_{CR}=1500\times 3\times 10^{-6}$ to have a short $\lambda$
since it is numerically difficult to treat such a long penetration depth.
We have confirmed that our main conclusions are essentially independent of the numerical value of $\varepsilon_{CR}$
by repeating the same calculations for several different values of $\varepsilon_{CR}$.

Although one of the two possible chiralities will be chosen by the magnetic flux density arising from $\tilde{M}^{\rm spin}$,
the chirality selection is a subtle issue, since there are several competing effects,
Landau diamagnetism, Meissner effect, orbital magnetization of the Cooper pair field~\cite{pap:Mineev2002,book:MineevSamokhin1999}, 
and chiral SC fluctuations~\cite{pap:Sumiyoshi2014}.
All these effects originate from Lorentz force on electron motions, while Zeeman effect on electron spins does not influence
chiralities of Cooper pairs although it favors a particular orientation of $d$-vector in spin-space under a given magnetic flux density~\cite{book:MineevSamokhin1999}.
The above first two contributions are diamagnetic and common to all the superconductors, 
while the latter two are characteristic to superconductors with broken time-reversal symmetry.
The orbital magnetization of the Cooper pair field $M^{\rm OP}_i\propto \epsilon_{ijk}d_{\mu j}^{\ast}d_{\mu k}$ 
depends on details of the system such as the density of states around the Fermi energy
and lattice symmetries,
and is usually negligibly small with the prefactor $(\Delta/\varepsilon_F)^2\ll 1$ within weak-coupling approximations~\cite{book:MineevSamokhin1999,pap:Furusaki2001}.
($d_{\mu j}$ is the Cooper pair field in a continuum system.)
On the other hand, coupling to the chiral SC fluctuations is paramagnetic and some corresponding coupling would be present also at low temperatures
well below superconducting transition temperatures~\cite{pap:Sumiyoshi2014}.
In the numerical calculations, if we put an initial configuration $\Delta^x_i=0,\Delta^y_i=$(position-independent tiny constant),
$\Delta^x_i$ with a SC domain wall is automatically induced when iteratively solving the gap equation, creating a chiral SC domain wall.
In this case, the orbital magnetization due to the surface current without Meissner effect is anti-parallel to $B_{iz}$,
which means that the magnetic response of chiral SC is diamagnetic. 
However, if we put an initial configuration of $\Delta^{x/y}_i$ so that the resulting magnetic response becomes paramagnetic,
we find that the paramagnetic solutions have slightly lower energies than the diamagnetic solutions.
These complicated behaviors may be due to the above mentioned competing effects.
In the following numerical calculations, we consider only the lower energy solutions of $\Delta$, i.e. paramagnetic solutions.
It is noted that, even if we use diamagnetic solutions, our main conclusions on the stability of SC domain walls and 
the resulting FF-like state essentially holds true.
Although we focus only on the Meissner state in the present study,
detailed structures of the gap functions around vortices will become important for the chirality selction
in the possible spontaneous vortex state
~\cite{pap:Ichioka2002,pap:Ichioka2005}.

\section{results}
For the moment, in order to understand the physics step by step, 
we do not take into account $\tilde{A}$ and $\tilde{B}$.
Since a chiral SC domain is not induced without $\tilde{B}$, we introduce an initial configuration of $\Delta_i$
with a SC domain which is consistent with the solutions under $\tilde{B}\neq 0$.
Inclusion of $\tilde{A}$ and $\tilde{B}$ leads to Meissner effect, and it will be discussed later.
The parameters are fixed as $g=5t$, filling $n=0.5$ and $h^{\rm ex}_0=0.1t, d_{\rm DW}=a$ for which 
the superconducting transition temperature is evaluated to be $k_BT_{\rm sc}\simeq 0.13t$.
We can estimate the coherence length of the SC at $k_BT=0.001t$ to be $\xi_0\simeq 6a$ from the calculated spatial profile of $\Delta_i$ 
near the surface/DW fitted by tanh$(x/\xi_0)$. 
Although the value of $\xi_0$ would be too small compared to the real values in chiral superconductors such as UCoGe,
a large coupling constant $g=5t$ simplifies our discussions.
Our main conclusions are essentially independent of absolute values of $\Delta_i$. 
In the following, we fix $k_BT=0.001t$ if not specified.

\subsection{$p_x-ip_y/-p_x-ip_y$ domains}
When a chiral superconducting DW is formed along the $yz$-plane by some reasons,
there will be two locally stable structures according to the previous Ginzburg-Landau (GL) study on $^3$He-A phase~\cite{pap:SalomaVolovik1989}.
One possibility is that $\Delta^x_i$ changes its sign at the domain boundary,
while $\Delta^y_i$ remains basically constant in space.
We call this domain structure ``odd-domain wall'' in this study since the $x$-odd component $\Delta^x_i$ changes sign at $x=x_{DW}$.
According to the previous study on $^3$He-A phase,
such a SC domain wall is metastable when it lies on the $yz$-plane.
More stable domain structure with the lowest DW energy is the gap function where
$\Delta^x_i$ remains nearly constant while $\Delta^y_i$ changes sign.
This domain structure is called ``even-domain wall'' in this study since the $x$-even component $\Delta^y_i$ changes sign at $x=x_{DW}$.
The relative stability of the even-DW is not so trivial in the present study, 
since there exist ``unpaired electrons'' only in the even-DW case which are not captured in GL theories 
and they might possibly destabilize the even-DW. 
Here, we firstly consider the former metastable case ($p_x-ip_y/-p_x-ip_y$ odd-DW) for completeness,
although much has been known for it~\cite{book:Volovik2003,pap:Mizushima_review2016}.
We will discuss the latter more stable
case ($p_x-ip_y/p_x+ip_y$ even-DW) later.
As mentioned before, we firstly do not take into account the magnetic flux density by setting $\tilde{B}_z=0$ for understanding the physics step by step, 
and Meissner effect will be discussed later.

\begin{figure}[htbp]
\begin{tabular}{cc}
\begin{minipage}{0.5\hsize}
\begin{center}
\includegraphics[width=\hsize,height=0.6\hsize]{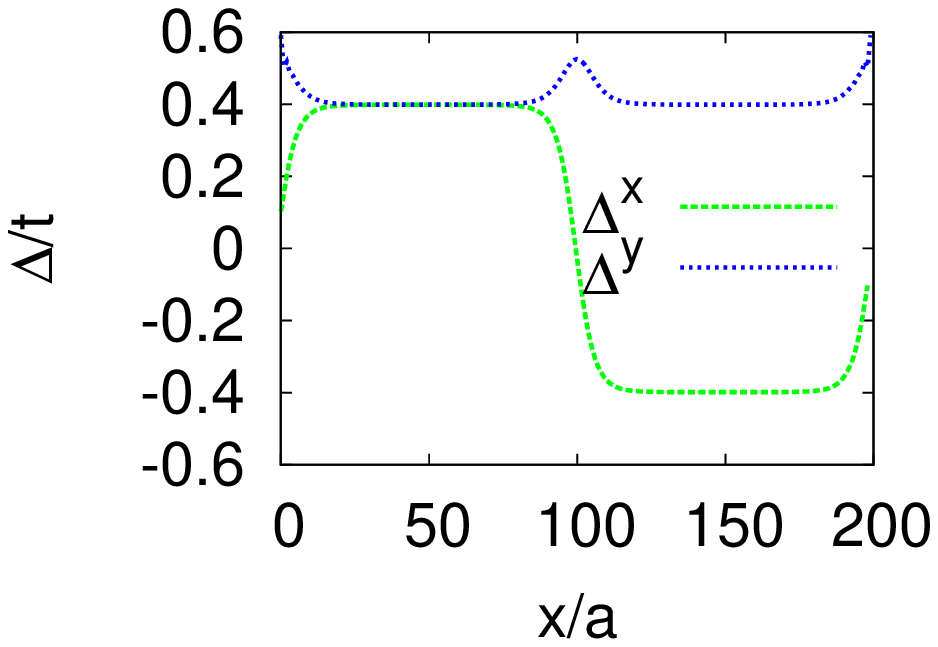}
\end{center}
\end{minipage}
\begin{minipage}{0.5\hsize}
\begin{center}
\includegraphics[width=\hsize,height=0.6\hsize]{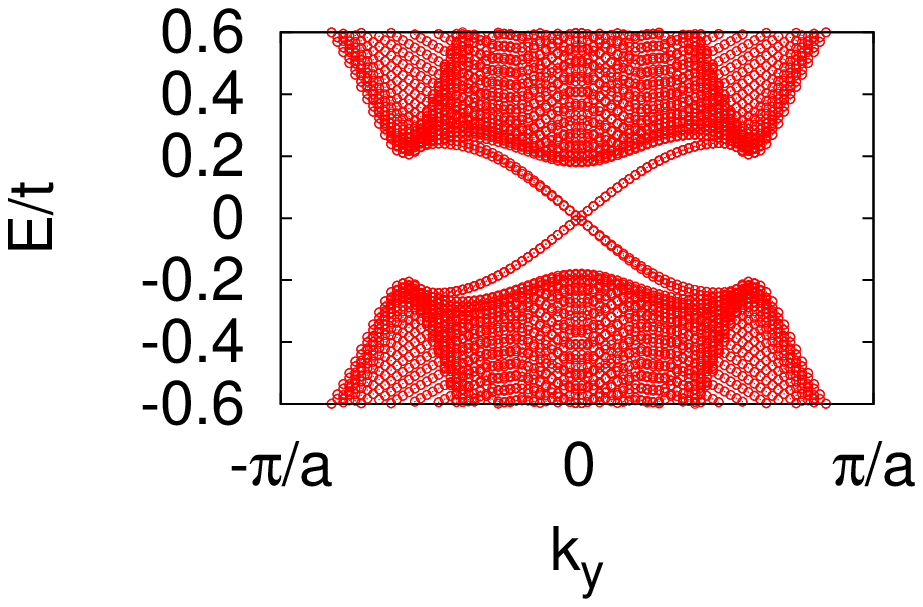}
\end{center}
\end{minipage}
\end{tabular}
\caption{(Left panel) Gap functions for the odd-DW case at $(g,k_BT,n)=(5t,0.001t,0.5)$.
The FM domain wall parameters are $(h^{\rm ex}_0,d_{\rm DW})=(0.1t,a)$.
(Right panel) Spectrum of the BdG Hamiltonian for the odd-DW case.}
\label{fig:Delta_odd}
\end{figure}

We show the gap function averaged over spins, $\Delta^{x/y}_i={\rm Re}\sum_{\sigma}\Delta^{x/y}_{i\sigma}/2$ in Fig. \ref{fig:Delta_odd},
where a chiral SC odd-domain structure is clearly seen.
It is noted that, since the chiral SC domains are stabilized by the FM domains, their spatial positions 
coincide each other. Therefore, the former could be easily identified in experiments by directly looking at the latter. 
This is an advantage of a charged ferromagnetic SC compared to the neutral paramagnetic $^3$He where identification of a superfluid domain is very difficult.

\begin{figure}[htbp]
\begin{center}
\includegraphics[width=0.8\hsize,height=0.5\hsize]{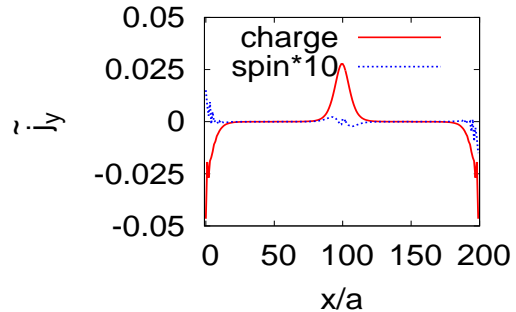}
\end{center}
\caption{Charge and spin currents without Meissner effect for the odd-DW case. 
}
\label{fig:j_odd}
\end{figure}

\begin{figure}[htbp]
\begin{center}
\vspace{0.5cm}
\includegraphics[width=0.7\hsize,height=0.2\hsize]{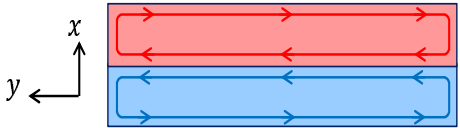}
\end{center}
\caption{The charge current profile for the odd-DW case in the system with boundaries. }
\label{fig:film_odd}
\end{figure}

We also show quasi-particle spectrum in Fig.~\ref{fig:Delta_odd} 
where there are two-hold degenerate surface/DW modes for each spin with 
the one-dimensional Fermi wavenumber $k_{F}^{\rm surf/DW}=0$.
The chirality of the superconductivity leads to 
surface/DW charge current $\tilde{j}_{i,y}$ and spin current 
$\tilde{j}_{i,y}^{\rm spin}$ as shown in Fig.~\ref{fig:j_odd}.
The charge current at a surface flows in the opposite direction to that of the DW current
in the same chiral SC domain as one would naively expect.
Because of this ``cancellation'', the net current along the $y$-direction in the whole sample or in a single domain vanishes.
If we consider an open boundary condition for the $y$-direction which is more realistic for an anisotropic sample,
the surface current will be smoothly connected to the DW current making a large circular current profile in 
the $xy$-plane
as shown in Fig.~\ref{fig:film_odd}.
On the other hand, the magnitude of the spin current at the surface is larger than that at the DW as seen Fig.~\ref{fig:j_odd}.
Therefore, the total spin current in one domain is non-zero in contrast to the charge current,
although the net spin current along the whole sample vanishes.

\begin{figure}[htbp]
\begin{tabular}{cc}
\begin{minipage}{0.5\hsize}
\begin{center}
\includegraphics[width=\hsize,height=0.6\hsize]{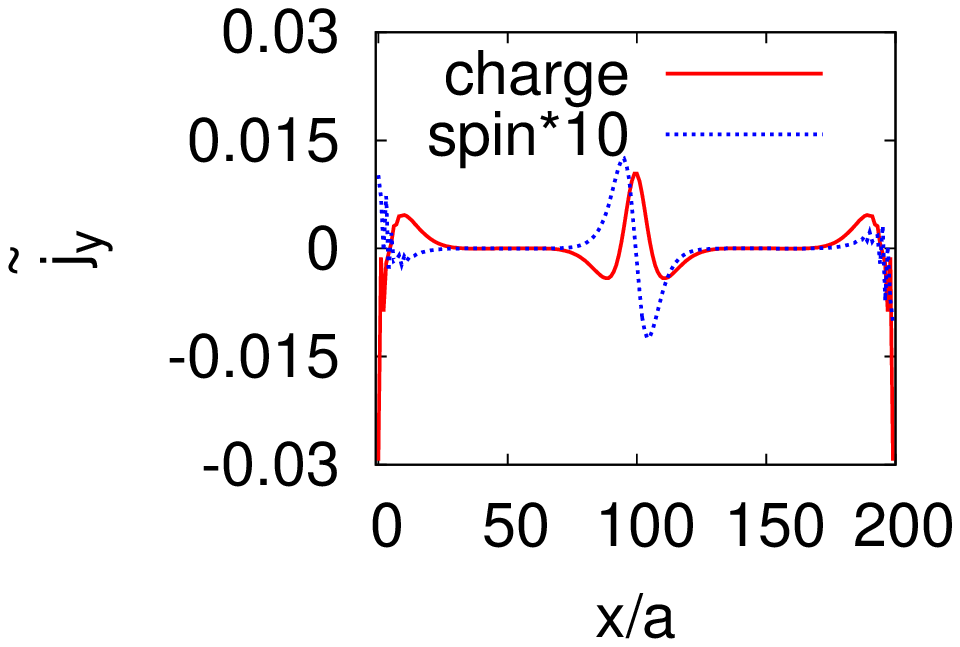}
\end{center}
\end{minipage}
\begin{minipage}{0.5\hsize}
\begin{center}
\includegraphics[width=\hsize,height=0.6\hsize]{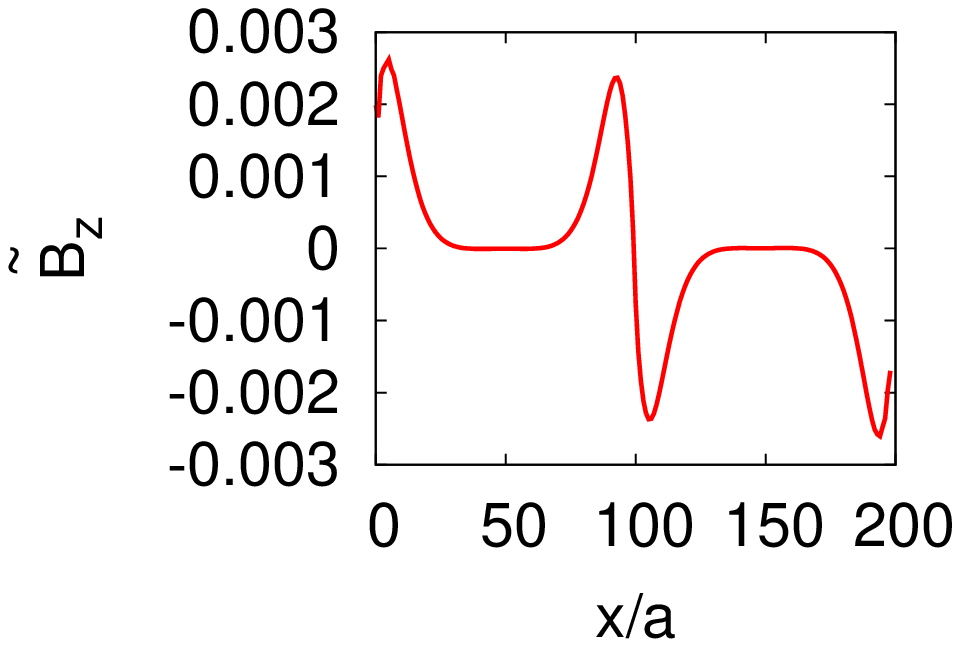}
\end{center}
\end{minipage}
\end{tabular}
\caption{(Left panel) Charge and currents with Meissner effect for the odd-DW case.
(Right panel) Corresponding magnetic flux density.}
\label{fig:j_oddM}
\end{figure}

We now discuss Meissner effect in this state by solving Maxwell equation together with the gap equation.
A computed current profile is shown in Fig.~\ref{fig:j_oddM}.
The Meissner effect simply induces screening current, and the resulting current density
is oscillating near the surface and DW~\cite{pap:Furusaki2001,pap:Ashby2009}.
Because of this oscillation, the net surface/DW current vanishes at a surface/DW,
which leads to vanishing $\tilde{B}_{iz}$ in a central region of a domain as shown in Fig.~\ref{fig:j_oddM}.

\subsection{$p_x-ip_y/p_x+ip_y$ domains}
As mentioned before, the $p_x-ip_y/-p_x-ip_y$ odd-domain is metastable and 
the $p_x-ip_y/p_x+ip_y$ even-DW has a lower energy 
in the case of $^3$He-A phase within the GL theory where the unpaired electrons are not taken into account~\cite{pap:SalomaVolovik1989}.
If this holds true also in the present system, the latter domain will be stabilized.
As in the previous section, the paramagnetic solutions of $\Delta$ have lower energies than 
those of the diamagnetic solutions also for the odd-DW.
Therefore, we only consider the paramagnetic solutions which show paramagnetic responses to the magnetization induced $\tilde{B}_{iz}$.
Similarly to the odd-DW, we do not consider Meissner effect for a while to understand the physics step by step.

\subsubsection{gap function and spectrum}
\begin{figure}[htbp]
\begin{tabular}{cc}
\begin{minipage}{0.5\hsize}
\begin{center}
\includegraphics[width=\hsize,height=0.6\hsize]{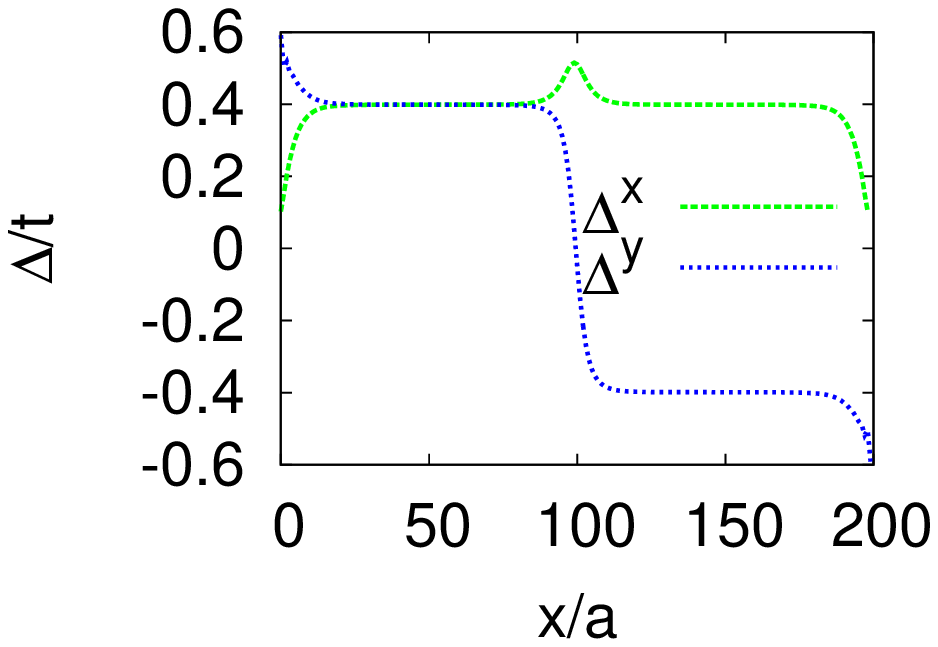}
\end{center}
\end{minipage}
\begin{minipage}{0.5\hsize}
\begin{center}
\includegraphics[width=\hsize,height=0.6\hsize]{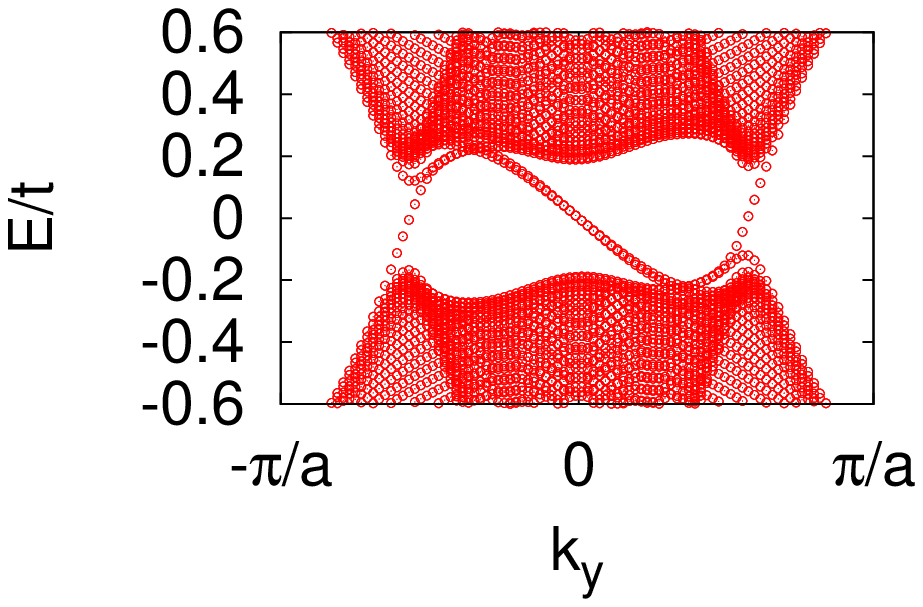}
\end{center}
\end{minipage}
\end{tabular}
\caption{(Left panel) Gap functions for the even-DW case.
(Right panel) Spectrum of the BdG Hamiltonian for the even-DW case.}
\label{fig:Delta_even}
\end{figure}

Figure \ref{fig:Delta_even} shows the gap functions and spectrum of BdG Hamiltonian.
The chiral domain is created as in the previous odd-DW case.
On the other hand, 
the quasi-particle spectrum looks complicated compared to the previous one (Fig. \ref{fig:Delta_odd}),
where the surface modes are almost unchanged while the DW modes are largely modified~\cite{pap:SilaevVolovik2012,pap:Tsutsumi2014,pap:Volovik2015}.
The DW modes cross zero energy at $k_y=0$ for the odd-DW, while the zero energy state is shifted to $k_y\simeq k_F$
in the even-DW case where $k_F$ is the Fermi wavenumber in the bulk~\cite{pap:Tsutsumi2014,pap:Volovik2015}.
The difference in the DW modes for the two DWs can be easily understood within a simple semi-classical discussion.
For simplicity, we here consider a continuum limit which is legitimate at low filling.
The one-dimensional ``Fermi wavenumber'' of the DW mode $k_F^{\rm DW}$ can be roughly evaluated by the vanishing quasi-particle energy,
\begin{align}
0&=[E(x,k_x,k_y)]^2\nonumber \\
&=\Bigl(\hbar^2\frac{k_x^2+k_y^2}{2m}-\mu\Bigr)^2+\Bigl(\Delta^x(x)\frac{k_x}{k_F}\Bigr)^2+\Bigl(\Delta^y(x)\frac{k_y}{k_F}\Bigr)^2,
\end{align}
where $k_x^2+k_y^2=k_F^2$ with $k_F=\sqrt{2m\mu}/\hbar$.
In the odd-DW case, $\Delta^x$ changes its sign at the DW and $\Delta^x(x=x_{DW})=0$ is satisfied as in Fig.~\ref{fig:Delta_odd},
which means that $[\Delta^y(x_{DW})k_y]^2=0$ is required, i.e. $k_F^{\rm DW}=0$.
In the even-DW case, however, $\Delta^y(x=x_{DW})=0$ holds as in Fig.~\ref{fig:Delta_even} 
and $[\Delta^x(x_{DW})k_x]^2=0$ requires $k_y^2=k_F^2$, i.e. $[k_F^{\rm DW}]^2=k_F^2$.
This explains the essential difference in the DW modes for the two domain structures. 

\begin{figure}[htbp]
\begin{center}
\includegraphics[width=\hsize,height=0.3\hsize]{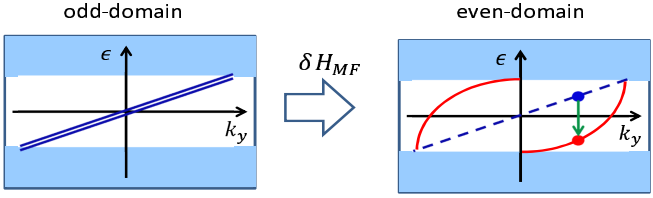}
\end{center}
\caption{Schematic picture of the spectral flow of the BdG eigenvalues. Only the DW modes are explicitly shown.
Some eigenvalues of the DW modes change signs when Hamiltonian is modified by $\delta H_{\rm MF}$ as indicated by the green arrow.
}
\label{fig:specflow}
\end{figure}

Generally, spectrum of surface/DW modes are neither protected by symmetry nor an intrinsic bulk property
but are sensitive to boundary conditions.
Topological arguments ensure only the existence of symmetry-protected surface/DW modes.
In the present system, the spectrum of the DW modes strongly depend on details of SC domain walls and
are modified by the change in Hamiltonian $\delta H_{\rm MF}=H^{\rm even}_{\rm MF}-H^{\rm odd}_{\rm MF}$
as shown in Fig.~\ref{fig:specflow}.
Some eigenvalues of the DW modes change signs by $\delta H$, which is called spectral flow of BdG eigenvalues
~\cite{book:Volovik2003,pap:NiemiSemenoff1986,pap:StoneGaitan1987,pap:Volovik2015,pap:Tada2015PRL,pap:Prem2017,pap:Huang2015,pap:Mizushima_review2016,com:specflow}.
The spectral flow is related to physical quantities in some models, and also to ground state wavefunctions
of mean field Hamiltonians in general, as will be discussed in the next section.

\subsubsection{current and unpaired electrons}
\begin{figure}[htbp]
\begin{center}
\includegraphics[width=0.8\hsize,height=0.5\hsize]{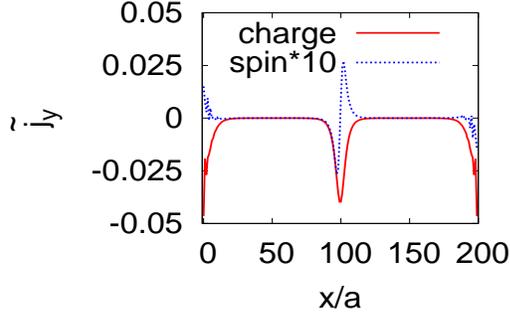}
\end{center}
\caption{Charge and spin currents without Meissner effect for the even-DW case.}
\label{fig:j_even}
\end{figure}

Although the $p_x-ip_y/p_x+ip_y$ even-DW is more stable, 
the charge current profile of this state is rather counter-intuitive as shown in Fig.~\ref{fig:j_even}.
While the surface current is essentially the same as that in the previous $p_x-ip_y/-p_x-ip_y$ odd-DW case,
the DW current flows in the opposite direction to the previous one.
Such reversal of the current direction has been discussed previously, and is possible in other systems
~\cite{pap:Tsutsumi2014,pap:Volovik2015,pap:Bouhon2014}.
A related issue has also been discussed before~\cite{pap:Wong2013,pap:Daido2017}.
The spin current at the domain wall is reversed similarly to the charge current and enhanced compared to
that of the previous odd-DW.

The counter-intuitive charge current can be attributed to the previously mentioned ``unpaired electrons" which arise from an implicit
pair breaking effect~\cite{pap:Tada2015PRL,pap:Prem2017}.
To see this, we again consider a continuum model for simplicity, which can be verified at low filling.
For such a case, the net current in the whole sample can be simply decomposed into the paramagnetic and diamagnetic parts,
$J_y=J_y^p+J_y^d$, where $J_y^p=\sum_{xk_{y}\sigma}e\hbar k_y/m c^{\dagger}_{xk_{y}\sigma}c_{xk_{y}\sigma}$
and $J_y^p=\sum_{xk_{y}\sigma}-e^2A_y(x)/(mc) c^{\dagger}_{xk_{y}\sigma}c_{xk_{y}\sigma}$.
Within the mean field approximation, the expectation value at zero temperature $\langle J_y^p\rangle_0$ is calculated as
\begin{align}
\langle J_y^p\rangle_0&= -\frac{1}{4}\sum_{k_y}\frac{e\hbar k_y}{m}\eta(k_y),\\
\eta(k_{y})&=\sum_n {\rm sign}[E_n(k_y)],
\end{align}
where $\{E_n(k_{y})\}$ are the eigenvalues of the BdG Hamiltonian ${\mathcal H}_{\rm BdG}$~\cite{pap:Tada2015PRL,pap:Prem2017,pap:Huang2015,pap:Mizushima_review2016}. 
$\eta$ is called spectral asymmetry and it is essentially determined by the Fermi wavenumber of the surface/DW modes,
because formations of surface/DW modes can cause spectral flow where some eigenvalues of ${\mathcal H}_{\rm BdG}$
change the signs depending on the model parameters, as shown in Fig.~\ref{fig:specflow}.
$\eta(k_y)=0$ everywhere in the odd-DW, while
$\eta(k_{y})$ is given by $\eta=+2 (-k_F<k_y<0), \eta=-2 (0<k_y<k_F)$ for each spin, and 
$\eta=0$ otherwise in the even-DW.
Because $J_y^p=J_y-J_y^d$ is satisfied, $\langle J_y^p\rangle_0=0$ simply means $\langle J_y\rangle_0=\langle J_y^d\rangle_0$,
which is usually expected for a full gap superconductor in the presence of a given vector potential.
On the other hand, a non-zero $\langle J_y^p\rangle_0$ or equivalently $\langle J_y\rangle_0\neq \langle J_y^d\rangle_0$
implies existence of a normal state (non-superconducting) components, i.e. unpaired electrons.

We also discuss finite temperature effects for the even-DW. 
Our numerical calculations show that the DW current is parallel to the surface current and 
the resulting net paramagnetic current is negative for all $0\leq T< T_{\rm sc}$ (not shown).
This can be explicitly shown within a quasi-classical calculation for the continuum model which linearizes the derivative,
$(k_F-i\partial_x)^2\simeq k_F^2-2ik_F\partial_x$~\cite{book:Volovik2003}.
It gives $E^{\rm surf}(k_y)=-\Delta k_y/k_F$ and $E^{\rm DW}(k_y)=\pm \Delta \sqrt{k_F^2-k_y^2}/k_F$ where
$\Delta$ is the gap amplitude in the bulk.
These spectrum are good approximations of our numerical results Fig.~\ref{fig:Delta_even} and also of 
the previous studies~\cite{pap:Tsutsumi2014,pap:SilaevVolovik2012,pap:Mizushima_review2016}.
The finite temperature effects on $\langle J_y^p\rangle$ is that (i) $\Delta(T)$ in $E_n(k_y)$ becomes smaller and 
(ii) the spectral asymmetry is modified to $\eta(k_y)=-\sum_n [2f(E_n(k_y))-1]$ with $f(E)=1/(e^{\beta E}+1)$ and $\beta=1/(k_BT)$.
Since contributions from the continuum spectrum $|E_n(k_y)|>\Delta$ will cancel between $+E$ and $-E$, they do not contribute to $\eta(k_y)$ at finite temperature,
which allows us to focus only on the remainig contributions from the surface and DW modes.
Then, it is easy to see that $\eta(k_y;T\geq 0)<0 (>0)$ at $0<k_y<k_F (-k_F<k_y<0)$ for the above spectrum $E^{\rm surf}$ and $E^{\rm DW}$,
although the magnitude $|\eta(k_y;T)|$ gets suppressed as $T$ increases. 
Therefore, $\langle J_y^p\rangle<0$ even at any finite temperature $0\leq T< T_{\rm sc}$,
which means that the DW current is parallel to the surface current.
On the other hand, the DW current within the GL theory is essentially determined by the SC chiralities only,
$j_y^{\rm GL}=n_s[(\delta_{yj}-l_yl_j/2)m_k\partial_jn_k+(\delta_{yj}/4-l_yl_j)(\nabla\times \vec{l})_j]$ with the conventional notation
$d_{\mu j}=d_{\mu}(m_j+in_j), \vec{l}=\vec{m}\times \vec{n}$ in the continuum model~\cite{book:Tsuneto1998},
and its direction is the same for both the odd-DW and even-DW.
The GL theory fails to describe the current reversal for the even-DW. 
This is because the current reversal is due to the hidden normal state contributions, 
and these unpaired electrons are not captured within the GL descriptions which focus only on superconducting contributions.

The unpaired electrons can be directly found in the ground state wavefunction of the mean field Hamiltonian
when $\langle J_y^p\rangle_0\neq 0$ arising from non-trivial $\eta(k_{y})\neq 0$. 
It can be shown that 
the ground state wavefunction is given by the tensor product of
\begin{align}
|{\rm MF}\rangle_{k_{y}}&=\prod_{n=1}^{n_1(k_{y})}c'^{\dagger}_{nk_{y}}\prod_{n=1}^{n_2(k_{y})}c'^{\dagger}_{n,-k_{y}}\nonumber\\
&\quad \times \exp[\sum_{n>n_1,n'>n_2}F_{nn'}(k_{y})c'^{\dagger}_{nk_{y}}c'^{\dagger}_{n',-k_{y}}]|0\rangle,
\end{align}
for each $\sigma$ ($\sigma$-index has been suppressed for simplicity),
where $|0\rangle$ is the vacuum of $c$-operators. 
$c'$-operators are given by 
superpositions of $c$-operators and $F_{nn'}(k_{y})$ can be calculated from the unitary matrix which
diagonalizes the BdG Hamiltonian~\cite{pap:Tada2015PRL,pap:Prem2017,pap:Lanbote1974}.
The exponential factor describes the paired electrons, while
the additional $\prod c'^{\dagger}\prod c'^{\dagger}$ corresponds to the unpaired electrons.
The number of the unpaired electrons for a given $k_{y}$ can be calculated from $\eta(k_{y})$;
$(n_1,n_2)=(1,0)$ for $(-k_F<k_y<0)$ while $(n_1,n_2)=(0,1)$ for $(0<k_y<k_F)$.
Therefore, these unpaired electrons carry momenta which are {\it opposite} to the given SC chirality,
which results in the reversal of the DW current from the one naively expected from the SC chirality.
We stress that these normal state components are not easily visible in the Hamiltonian Eq.\eqref{eq:H} or \eqref{eq:H_MF}
and are absent in the odd-DW case and uniform SC,
but they do exist in the present stable even-DW case.
These hidden contributions are important for the DW current induced by the momentum space topology.
In addition, they can contribute also to other quantities such as specific heat and NMR relaxation rate,
although the contributions would be sub-dominant in magnitude.
Furthermore, we will show in the next section that the existence of the unpaired electrons leads to
a FF-like state.

\subsubsection{stability compared to odd-DW}
\begin{figure}[htbp]
\begin{center}
\includegraphics[width=0.7\hsize,height=0.5\hsize]{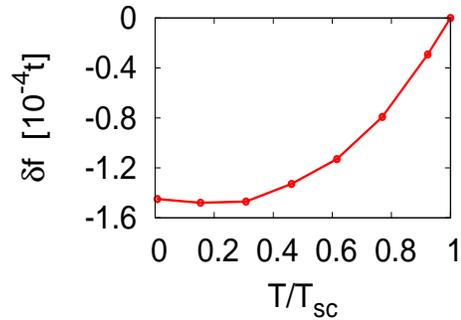}
\end{center}
\caption{Free energy difference between even-DW and odd-DW, $\delta f$, as a function of temperature $T$.}
\label{fig:df_T}
\end{figure}

Since formation of Cooper pairs usually reduces energy in superconductors, one may think that
unpaired electrons would increase energy and eventually destabilize the even-DW.
Indeed, in some simple examples, unpaired electrons can appear 
only in extreme parameter regions where pair breaking effects are so strong that the superconductivities become unstable~\cite{com:Tada}.
In the previous GL theory, it was shown that even-DWs are more stable than odd-DWs~\cite{pap:SalomaVolovik1989}.
However, since the unpaired electrons are not taken into account and temperature dependence has not been discussed in the previous GL study, 
stability of the even-DW at both zero and finite temperatures 
is non-trivial.
In order to examine stability of the even-DW structure, 
we compute the free energy density $\delta f$ measured from that of 
the odd-DW system,
\begin{align}
\delta f&=f_{\rm even}-f_{\rm odd},\\
\label{eq:df_eo}
f&=-\frac{1}{L_xL_y}k_BT\log {\rm tr}(e^{-\beta H_{\rm MF}}),
\end{align}
where $\beta$ is the inverse temperature.
As shown in Fig.~\ref{fig:df_T}, the even-DW is stable up to all the temperatures below $T_{\rm sc}$.
The stability of the even-DW will leads to a stable FF-like state at the same temperature range, $0\leq T\leq T_{\rm sc}$
as will be discussed later.
It is noted that $|\delta f|$ becomes smaller (larger) as the thickness of the $x$-direction is increased (decreased),
since the free energy difference is proportional to the DW area $L_y$, i.e. $\delta f\sim O(1/L_x)$.

Our numerical calculations would imply that DW energies are mainly determined
by spatial profiles of $\Delta^{x/y}_i$
and quasi-particle contributions do not change the stability of the even-DW,
although their contributions to the energy would be of the same order $O(L_y)$.
Indeed, by comparing $\Delta^{}_i$ for the two DWs (Figs.~\ref{fig:Delta_odd} and ~\ref{fig:Delta_even}),
we see that $\Delta^y$ rather sharply changes at the even-DW
while $\Delta^x$ is suppressed in a wider region near the odd-DW, which results in 
a relatively strong pair-breaking effect of the gap function in the odd-DW case.
The different spatial profiles of the gap functions at the two DWs 
come from the fact that $\Delta^y$ is an even-parity gap function in the $x$-direction
while $\Delta^x$ is odd in $x$, which can be captured also in Ginzburg-Landau formalism as in the previous study~\cite{pap:SalomaVolovik1989}.
Even parity gap functions are generally more stable against the surface pair breaking effect on gap functions than odd parity ones.
The important point is that this mechanism dominates $\delta f$ even when contributions from the unpaired electrons are included
within the BdG approach which are of the same order, $O(L_y)$.

\subsubsection{Fulde-Ferrell-like state}
The net charge current along the $y$-direction is non-zero in the present gap function.
Generally such a current carrying state is not forbidden for example in three-dimensional toroidal geometry
which may be mimicked by a periodic boundary condition~\cite{pap:TadaKoma1,pap:Tada2015}.
However, it would be an unphysical state in a realistic anisotropic system with open boundaries, 
since there is no current profile which consistently connects the unidirectional surface and DW currents
with keeping the continuity equation $\nabla\cdot j=0$.
Besides, the present model might be physically regarded as a composition of one-dimensional ribbons along the $y$-direction 
with width $L_x$ stacking in the $z$-direction.
It is known that a current carrying state can never be a ground/equillibrium  state in the thermodynamic limit in one-dimension~\cite{pap:Bohm1949},
which may imply absence of a current carrying state in the present system, too.
Indeed, if we simply apply a one-dimensional Bloch-like argument, 
we can show that the current carrying state is unstable and a Fulde-Ferrell-like state with
vanishing net current is more stable, which can be confirmed at least within mean field approximations.

$\quad$

\noindent
{\it Optimization of $Q_y$ and stability}

\begin{figure}[htbp]
\begin{tabular}{cc}
\begin{minipage}{0.5\hsize}
\begin{center}
\includegraphics[width=0.9\hsize,height=0.6\hsize]{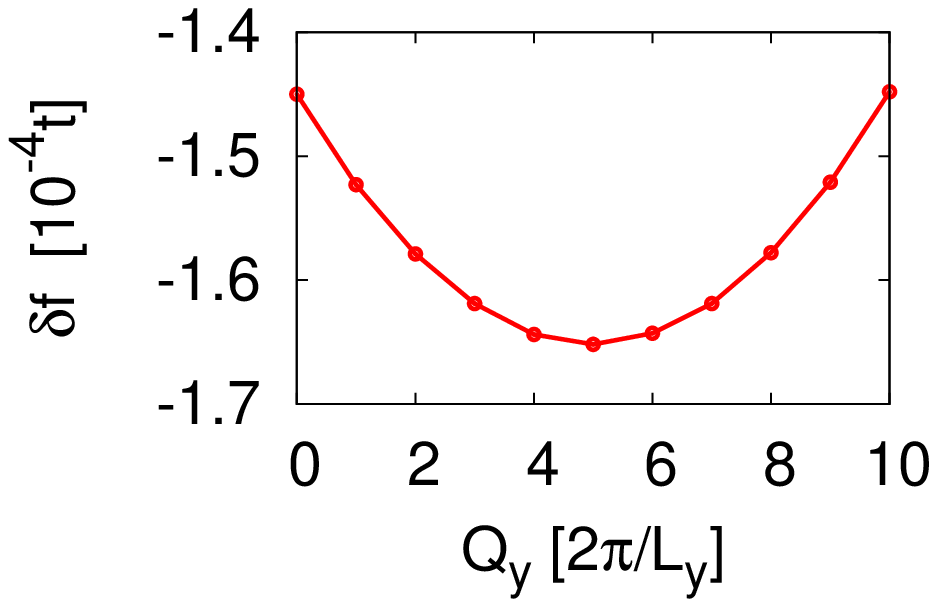}
\end{center}
\end{minipage}
\begin{minipage}{0.5\hsize}
\begin{center}
\includegraphics[width=\hsize,height=0.6\hsize]{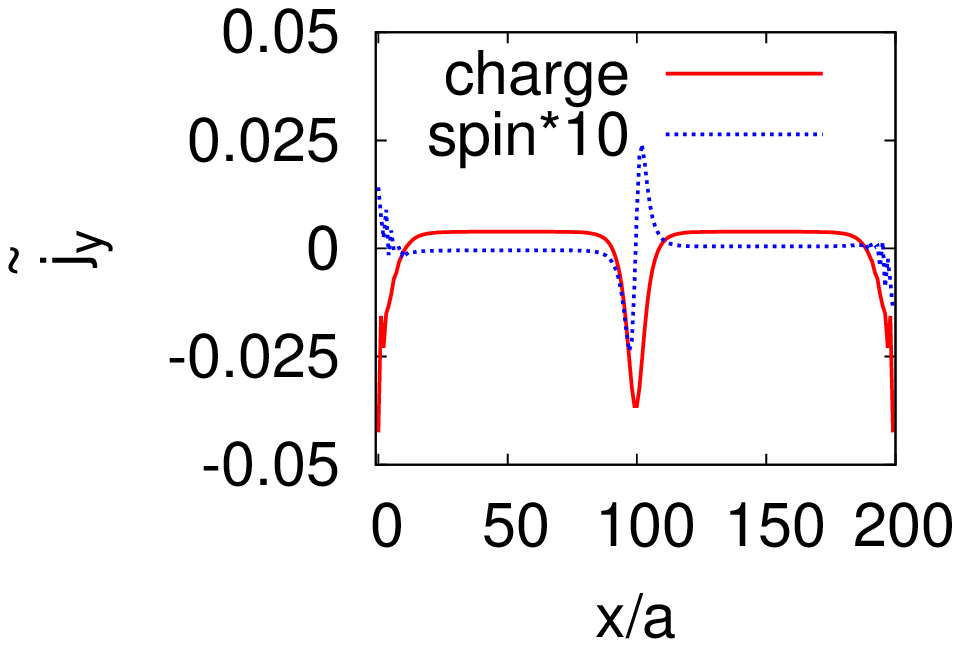}
\end{center}
\end{minipage}
\end{tabular}
\caption{(Left panel) Free energy difference as a function of the modulation $Q_y$.
(Right panel) Charge and spin currents without Meissner effect for the FF-like state.}
\label{fig:df_Q}
\end{figure}

\begin{figure}[htbp]
\begin{center}
\includegraphics[width=0.7\hsize,height=0.2\hsize]{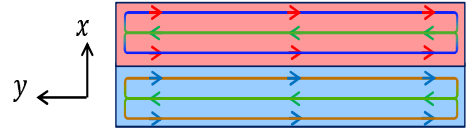}
\end{center}
\caption{The charge current profile for the FF-like state in the system with boundaries. 
The surface and DW currents flow in the same direction, while the FF-like supercurrent flows in the opposite direction.}
\label{fig:film_FF}
\end{figure}

We consider a FF-like stripe state with 
\begin{align}
g\langle c_{i\sigma}c_{i+\hat{x},\sigma}\rangle&=-i\Delta^x_{i\sigma} e^{iQ_yy}, \\
g\langle c_{i\sigma}c_{i+\hat{y},\sigma}\rangle&=\Delta^y_{i\sigma} e^{iQ_y(y+a/2)}, 
\end{align}
where $Q_y=2\pi n_y/L_y (n_y=$integer) is the center of mass momentum of a Cooper pair
and $\Delta^{x/y}_{i\sigma}$ depend only on $x$.
Here, we have assumed the same oscillation periods for $\Delta^{x}$ and $\Delta^y$, $Q_{xy}=Q_{yy}\equiv Q_y$,
since $Q_{xy}\neq Q_{yy}$ would lead to additional chiral SC domain walls in the $y$-direction,
which increases energy.
The negligibly small factor $e^{iQ_ya/2}$ has been introduced for the later convenience.
We minimize the free energy density $\delta f(Q_y)$ Eq.~\eqref{eq:df_eo} with respect to $Q_y$
as shown in Fig.~\ref{fig:df_Q}.
Difference in the free energy density is small because it arises from the energy cost 
proportional to the area $L_y$ but not to the volume $L_xL_y$.
The resulting state with the minimum energy has vanishing net current, because the sum of the DW and surface current
cancels with the FF-like current flowing in the bulk region with the optimal $Q_y=Q_y^{\ast}\sim O(1/L_x)$, as shown in Fig.~\ref{fig:df_Q}.
If we consider an open boundary condition for $y$-direction, we can now naively expect that 
corresponding spatial configuration of the current will show a consistent profile as in Fig. ~\ref{fig:film_FF}.

The coincidence between the values of $Q_y$ for the vanishing total current and free energy minimum is of course not accidental,
and is related to Bloch's theorem on the absence of bulk current at equilibrium~\cite{pap:TadaKoma1,pap:Tada2015,pap:Bohm1949}.
Although the Bloch's theorem can also be applied to superconductors, 
we should be careful when applying it to a mean field Hamiltonian where global U(1) symmetry is explicitly broken.
We introduce the {\it global} twist operator $U=\exp(i\delta Q_y P_y)$ with $P_y=\sum_{i\sigma}yn_{i\sigma}$
where the summation is over all the sites and $\delta Q_y=2\pi/L_y$.
The anihilation operators are transformed as $Uc_{j\sigma}U^{\dagger}=e^{i\delta Q_yy}c_{j\sigma}$ under the twist.
Although one needs to use a {\it local} twist operator for a rigorous discussion on an infinite volume system, 
the global operator simplifies our discussion.
The free energy difference between the $Q_y+\delta Q_y'$-state with $\delta Q_y'=2\delta Q_y$ and $Q_y$-state within the mean field approximation
is given by
\begin{align}
\beta \delta F&=-\log {\rm tr}(e^{-\beta H_{\rm MF}(Q_y+\delta Q_y')})+\log {\rm tr}(e^{-\beta H_{\rm MF}(Q_y)})\nonumber \\
&=-\log {\rm tr}(e^{-\beta UH_{\rm MF}(Q_y+\delta Q_y')U^{\dagger}})+\log {\rm tr}(e^{-\beta H_{\rm MF}(Q_y)})\nonumber \\
&=\beta(\hbar/e){\rm tr}(J_ye^{-\beta H_{\rm MF}(Q_y)})\delta Q_y'+O(\delta Q_y'^2),
\label{eq:dF_Q}
\end{align}
where $J_y=\sum_i j_{i,y}$ is the total current operator.
Here, we have assumed that the mean field solutions depend weakly on $Q_y$,
i.e. $\Delta^{x/y}_i(Q_y+\delta Q_y')=\Delta^{x/y}_i(Q_y)+O(\delta Q_y'^2)$.
The first term is estimated as $O(L_y\delta Q_y)=O(1)$ since the current density is localized
near the surface/boundary or vanishingly small in the bulk as implied by Bloch's theorem~\cite{pap:TadaKoma1,pap:Tada2015}, 
while the second term would be $O(L_xL_y\delta Q_y^2)=O(L_x/L_y)$.
In general two- or three-dimensional systems, the first term is not small compared to the second term, 
and a current carrying equilibrium state is not excluded based on this argument.
However, 
since $L_x\ll L_y$ is satisfied and the present system is more like one-dimensional, 
the first term dominates and an energy minimum $\delta F\sim O(L_x)$ corresponds to
a vanishing total current $\langle J_y\rangle \simeq 0\times L_y$.

As seen in Fig.~\ref{fig:df_T}, the even-DW is stable at all $0\leq T\leq T_{\rm sc}$ (Fig.~\ref{fig:df_T}),
which means that the FF-like state has a lower free energy in the same temperature region according to Eq. \eqref{eq:dF_Q}.
Therefore, the FF-like state can be realized at any temperature below $T_{\rm sc}$.
We note that similar FF-like states have been proposed in different models~\cite{pap:Kusama1999,pap:Vorontsov2009,
pap:Hachiya2013,pap:Mizushima_review2016,pap:Vorontsov2016,pap:Higashitani2015}.
Generally, these states can be stabilized only at low temperatures, 
while the robust stability of our FF-like state up to $T_{\rm sc}$ is characteristic to the FM chiral superconductors.
Besides, the thickness of the present model is much larger than the superconducting coherence length, $L_x\sim 30$-40$\xi_0$
and the emergence of the FF-like state itself is robust against the thickness, which
is in sharp contrast to the previous studies where FF-like states can be stable only for $L_x\sim \xi_0$ at finite temperatures
~\cite{pap:Vorontsov2009,pap:Hachiya2013,pap:Mizushima_review2016,pap:Vorontsov2016,pap:Higashitani2015}.
Therefore, our results are basically applicable not only to thin films but also to relatively large anisotropic samples,
although the modulation $Q_y^{\ast}$ becomes smaller in the latter.
In addition, 
the previously proposed FF-like states are sensitive to surface conditions~\cite{pap:Higashitani2015}.
On the other hand, the present FF-like state is stable as long as the even-DW is realized by the FM domain, and the surface/DW current 
would not be strongly changed e.g. by surface roughness~\cite{pap:Ashby2009,pap:Nagato2011}.

$\quad$

\noindent
{\it Spectral asymmetry and ground state wavefunction}

In the presence of the FF modulation $Q_y$, the paramagnetic current operator does not commute with 
the mean field Hamiltonian $H_{\rm MF}$ in the continuum limit.
Instead, the combination ${\mathcal J}_y^p\equiv J^p_y-Q_y N/2$ now commutes with $H_{\rm MF}$,
where $N$ is the total number operator $N=\sum_{xk_y\sigma}c^{\dagger}_{xk_y\sigma}c_{xk_y\sigma}$.
Therefore, $\langle {\mathcal J}_y^p\rangle_0$ can be written in terms of the spectral asymmetry,
\begin{align}
\langle {\mathcal J}_y^p\rangle_0&= -\frac{1}{4}\sum_{k_y}\frac{e\hbar}{m}(k_y+Q_y/2)\eta(k_y).
\end{align}
This quantity gives a deviation of the paramagnetic current from the naively expected value $(e\hbar/m)Q_y\langle N\rangle_0/2$
in the FF-like state, and a non-zero  $\langle {\mathcal J}_y^p\rangle_0$ implies existence of normal state components, 
i.e. unpaired electrons~\cite{com:Tada}.
Since $\eta$ is only slightly changed by $Q_y^{\ast}\sim O(1/L_x)$, 
$\langle {\mathcal J}_y^p\rangle_0$ is non-zero and the unpaired electrons still exist in the FF-like state.
Indeed, the ground state wavefunction of the mean field Hamiltonian now contains
\begin{align}
|{\rm MF}\rangle_{k_{y}}&=\prod_{n=1}^{n_1^{Q_y}}c'^{\dagger}_{n,k_{y}+Q_y}\prod_{n=1}^{n_2^{Q_y}}c'^{\dagger}_{n,-k_{y}}\nonumber\\
&\quad \times \exp[\sum_{n>n_1,n'>n_2}F^{Q_y}_{nn'}(k_{y})c'^{\dagger}_{n,k_{y}+Q_y}c'^{\dagger}_{n',-k_{y}}]|0\rangle,
\end{align}
where all the parameters can be calculated from the unitary matrix which diagonalizes the BdG Hamiltonian.
Since the reversal of the DW current is due to the unpaired electrons,
the emergence of the FF-like state is also a direct result of such hidden components.
Although impacts of unpaired electrons on physical quantities such as edge currents have been discussed previously~\cite{pap:Tada2015PRL,pap:Prem2017},
the present study is the first example where the unpaired electrons affect 
the underlying gap functions from which they arise.
In addition, it is considered that the existence of the unpaired electrons and resulting FF-like state
are general properties in chiral superconductors/superfluids in the presence of chiral domains.
UCoGe could provide a good platform to study such physics.

$\quad$

\noindent
{\it Meissner effect}

\begin{figure}[htbp]
\begin{tabular}{cc}
\begin{minipage}{0.5\hsize}
\begin{center}
\includegraphics[width=\hsize,height=0.6\hsize]{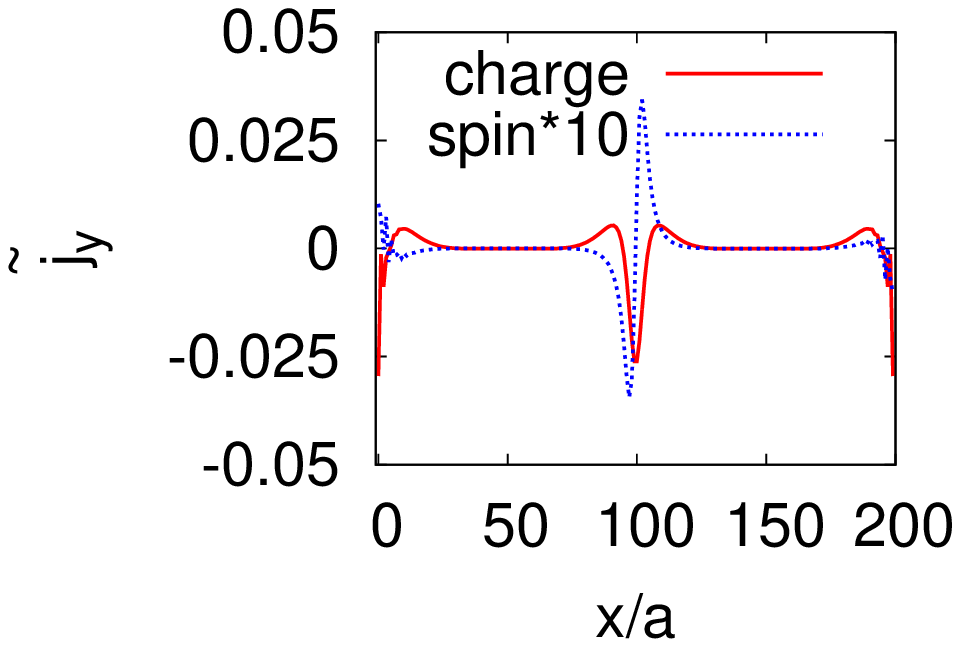}
\end{center}
\end{minipage}
\begin{minipage}{0.5\hsize}
\begin{center}
\includegraphics[width=\hsize,height=0.6\hsize]{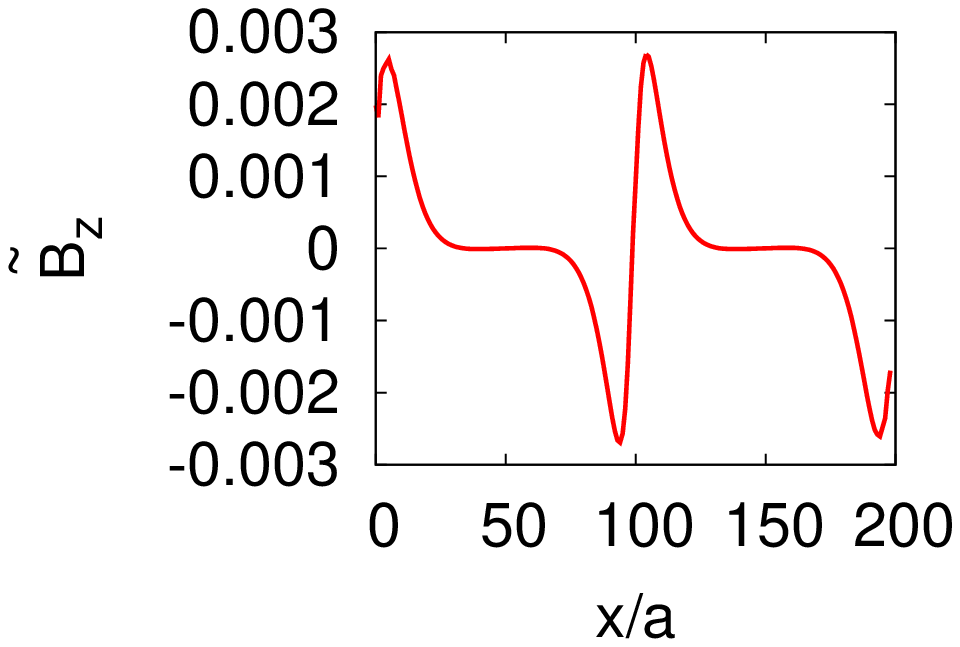}
\end{center}
\end{minipage}
\end{tabular}
\caption{(Left panel) Charge and spin currents with Meissner effect for the FF-like state.
(Right panel) Corresponding magnetic flux density.}
\label{fig:j_FFM}
\end{figure}

The current profile without Meissner effect now satisfies the continuity equation $\nabla\cdot j=0$ when we consider an open boundary
in the $y$-direction.
Such a current can be naturally screened by Meissner current if we solve 
the Maxwell equation self consistently. We show calculation results in Fig.~\ref{fig:j_FFM}.
The current in the bulk region is screened to be zero as in ordinary superconductors, in addition to the screening of the surface/DW current.
However, we note that it is also possible to have Meissner screening for example in $Q_y=0$ even-DW
and the resulting net current vanishes in our numerical calculations.
Such states have higher fermionic energies than the above state with the optimal modulation $Q_y^{\ast}$~\cite{com:Bterm}.

The unique charge current profile leads to a characteristic behavior of magnetic flux density as shown in Fig.~\ref{fig:j_FFM}.
Interestingly, the magnetic flux density $B_{iz}$ near the DW points to the opposite direction from the one
in the non-superconducting FM state.
We note that such reversal of magnetic flux density is possible only at low temperatures,
and magnetic flux density is smoothly changed as the system is cooled from a non-superconducting FM state.
An experimental observation of the reversed magnetic flux density would directly imply
the FF-like state.  
It is also interesting to see a possible feedback effect on FM domain walls.
Since the direction of the magnetic flux density around a domain wall is opposite to that in the non-superconducting state,
the FM domain walls will become thicker due to the Zeeman coupling.
As long as this Zeeman effect is not so strong and the initial FM domain configuration is not changed, the main results presented in this study should remain true. 
On the other hand, such an effect is not expected in the odd-DW case since $B_z$ is simply parallel to that in the non-superconducting state.
Similarly, feedback effects on FM properties would be weak in non-topological superconductors in the Meissner phases,
because the magnetic flux density will be qualitatively similar to that in odd-DW case (Fig.~\ref{fig:j_oddM}).
As mentioned in Sec.I, feedback effects could be discussed within a full self-consistent calculation for both FM and SC orders,
and are left for a future study.

$\quad$

\noindent
{\it Multi-domain structure}

\begin{figure}[htbp]
\begin{center}
\includegraphics[width=0.9\hsize,height=0.3\hsize]{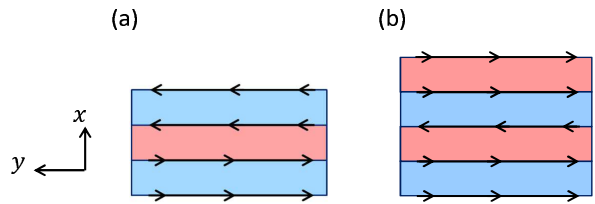}
\end{center}
\caption{Multi-domain structures. The number of DWs is even in (a) and odd in (b).
The arrows represent surface/DW currents when all the domain walls are even-DWs. 
The FF-like supercurrent is not shown.}
\label{fig:multidomain}
\end{figure}

Finally, as a straightforward extension of the above results, 
we consider a sequence of FM domains as in Fig.~\ref{fig:multidomain}.
In this case, the even-DW is favored at each FM domain wall, and each DW current is reversed from the naively expected ones
while only the surface currents are unchanged.
If the number of DWs is even, the DW currents completely cancel and also two surface currents cancel each other,
resulting in vanishing net current.
If the number of DWs is odd, almost all the DW currents cancel and the magnitude of the net current will be 
the sum of single DW current and two surface currents.
Therefore, the FF-like state can be stable only in the latter case, and the optimal modulation $Q_y^{\ast}$ is of the order $O(1/L_x)$
as in the single DW system discussed above.
On the other hand,
possible contributions from the unpaired electrons to other quantities such as specific heat and NMR relaxation rate will simply add up
as the number of DWs increases.
Although they would be still sub-dominant in magnitude, a careful consideration might be required in experiments.

\section{summary and discussion}
We have studied general properties of the ferromagnetic chiral superconductors by using the specific model of UCoGe
in the Meissner phase,
with focusing on interplay between momentum space topology and real space structure of FM domains.
We confirmed that, within the mean field approximation, 
the chiral SC domains are naturally induced by the FM domains because of Lorentz force arising from
the FM order which favors one of the two possible SC chiralities.
Besides, it was found that the relative stability among the different SC domain walls discussed in the previous GL theory
without the unpaired electrons
holds true also in our system where the unpaired electrons are explicitly taken into account.
In the metastable SC domain state, the current profile is simply a global circulation 
along the surface and the DW.
On the other hand, the DW current is reversed in the more stable SC domain as in $^3$He-A phase,
and the surface current and DW current do not cancel.
Because of this non-cancellation, the FF-like modulated SC state with vanishing net current can be stabilized
for all the temperatures below the superconducting transition temperature.
In a microscopic point of view,
the emergence of the FF-like state can be attributed to the hidden normal state components, i.e. the unpaired electrons.
Our study is the first example where the unpaired electrons affect the underlying gap functions from which they arise.
These results hold true also in other chiral superconductors/superfluids in the presence of chiral domains.
UCoGe provides a good platform for studying such physics.

Although we have used the mean field approximations in the present study,
mean field approximations usually overestimate stabilities of ordered states.
Rigorously speaking, one- or two-dimensional systems do not exhibit superconductivity at finite temperature
and a net current carrying state in two- and three-dimensions is not forbidden at least by Bloch's theorem~\cite{pap:TadaKoma1,pap:Tada2015}.
This might imply that our results based on the mean field approximations and essentially one-dimensional Bloch-like argument
on the mean field Hamiltonian are not correct enough.
More rigorous discussions are needed to clarify this issue.
It is also interesting to study the fate of the proposed FF-like state when the spontaneous vortex state is taken into account,
corresponding to lower pressure regions in UCoGe.

\section*{acknowledgement}
We thank S. Fujimoto and T. Mizushima for valuable discussions.
This work was supported by JSPS/MEXT Grant-in-Aid for Scientific Research
(Grant No. 26800177 and No. 17K14333)
and by a Grant-in-Aid for
Program for Advancing Strategic International Networks to
Accelerate the Circulation of Talented Researchers (Grant No.
R2604) ``TopoNet.''


\begin{thebibliography}{55}
\bibitem [{com({\natexlab{c}})}]{com:Bterm}%
  \BibitemOpen
  \href@noop {} {} \bibinfo {note} {Here, the Maxwell term
  $B^2/8\pi$ is not taken into account and only the fermionic part of the
  free energy is considered. Although evaluation of the Maxwell term is subtle
  since the present value of the numerical factor $\varepsilon_{CR}$ which
  controls the penetration depth is not a realistic value, inclusion of the
  Maxwell term does not change our conculsion.}\BibitemShut {Stop}%
\end{thebibliography}

%

\end{document}